\documentclass[12pt,a4paper]{article}
\usepackage{physics}
\usepackage{amsfonts}
\usepackage{amsmath}
\usepackage{amssymb}
\usepackage{amsthm}
\usepackage{graphicx}
\usepackage{jheppub}
\usepackage{url}
\usepackage{todonotes}
\usepackage{bm}
\usepackage{xcolor}
\usepackage{caption}
\usepackage{subcaption}
\usepackage{lmodern}
\usepackage[section]{placeins}
\usepackage[utf8]{inputenc}
\usepackage[T1]{fontenc}
\definecolor{refkey}{gray}{0.45}
\definecolor{labelkey}{RGB}{155,48,48}
\hypersetup{
  colorlinks=true,
  citecolor=purple,
  linkcolor=blue,
  urlcolor=violet
 }
 
 \newcommand{\mi}{\mathrm{i}}

\newcommand{\bi}{\begin{itemize}}
\newcommand{\ei}{\end{itemize}}

\newcommand{\bea}{\begin{eqnarray}}
\newcommand{\eea}{\end{eqnarray}}
\newcommand{\be}{\begin{equation}}
\newcommand{\ee}{\end{equation}}
\newcommand{\ben}{\begin{eqnarray*}}
\newcommand{\een}{\end{eqnarray*}}
\newcommand{\bem}{\begin{pmatrix}}
\newcommand{\eem}{\end{pmatrix}}
\newcommand{\bl}{\begin{align}}
\newcommand{\el}{\end{align}}
\newcommand{\beg}{\begin{gather}}
\newcommand{\eeg}{\end{gather}}

\newcommand{\cF}{\mathcal{F}}

\newcommand{\cH}{\mathcal{H}}

\newcommand{\cN}{\mathcal{N}}

\newcommand{\IH}{\mathbb{H}}

\newcommand{\e}{\epsilon}

\newcommand{\TrH[1]}{ {\raise -.5em
                      \hbox{$\buildrel {\textstyle  {\rm Tr } }\over
{\scriptscriptstyle \cH _ {#1}}$}~}}
\newcommand{\res[1]}{{\raise -.5em 
\hbox{$\buildrel{\textstyle{\rm Res}}\over {\scriptscriptstyle {#1}}$}}}
\newcommand{\tends[1]}{{\raise -.5em 
\hbox{$\buildrel{\longrightarrow}\over {\scriptscriptstyle {#1}}$}}}

\def\dbend{\lower3.5pt\hbox{\manual\char127}}

\def\IL{\relax{\rm I\kern-.18em L}}
\def\IH{\relax{\rm I\kern-.18em H}}
\def\rlx{\relax\leavevmode}

\def\ZZ{\rlx\leavevmode\ifmmode\mathchoice{\hbox{\cmss Z\kern-.4em Z}}
 {\hbox{\cmss Z\kern-.4em Z}}{\lower.9pt\hbox{\cmsss Z\kern-.36em Z}}
 {\lower1.2pt\hbox{\cmsss Z\kern-.36em Z}}\else{\cmss Z\kern-.4em
 Z}\fi}

\title{\center \textmd{Entanglement Entropy in String Compactifications}}

\preprint{}
\author{ \center Atish  Dabholkar and} \author{Upamanyu Moitra}
\affiliation{\begin{center}
Abdus Salam International Centre for Theoretical Physics\\ Strada Costiera 11, Trieste 34151, Italy
\end{center}  }

\abstract{We consider $\mathbb{Z}_N$ orbifolds of Type-II compactifications to four and six dimensions on several Calabi-Yau manifolds in the orbifold limit with the aim to compute the entanglement entropy. The spectrum can contain tachyons in the doubly-twisted sectors which can lead to new infrared divergences for the partition function that are not present in the orbifolds of the uncompactified ten-dimensional theory. We show that  all tachyonic contributions in these models  admit a resummation and analytic continuation that yields finite entropy in the physical region $0 < N \leq 1$  just as in ten dimensions. 
\vspace{5mm}
}

\makeatletter
\gdef\@fpheader{ \\ }
\makeatother
\begin{document}

\maketitle

\section{Introduction}\label{sec-intro}

A method to compute entanglement entropy in string theory was proposed in \cite{Dabholkar:1994ai, Dabholkar:2001if, Dabholkar:2022mxo} using a stringy analog of the replica method.  One can consider $\mathbb{Z}_N$ orbifolds of two-dimensional Euclidean Rindler plane $\mathbb{R}^2$,  where $N$ is an odd positive integer. The world-sheet partition function $Z(N)$ of these orbifolds is computable using standard orbifold techniques in perturbation theory. In the Lorentzian continuation, the orbifolds are related to closed strings on the Rindler horizon.
The partition function can be interpreted as the logarithm of the trace $\Tr(\rho^{1/N})$ where $\rho$ is the reduced density matrix  for the observer in the right Rindler wedge in the Minkowski vacuum.   Given this data for all odd positive integers $N$, one can seek an analytic continuation of the partition function to the physical region $0 < N \leq 1$ whose uniqueness can be guaranteed if the conditions required by the Carlson theorem  \cite{Carlson1914, Boas:1954}
 are satisfied. If a sensible analytic continuation exists,  then one can compute what can be regarded as the stringy quantum gravitational analog of the R\'enyi and von Neumann entropy. 

The foremost motivation for attempting such a computation is the fact that  string theory is a consistent quantum theory of gravity that is ultraviolet finite. Entanglement entropy  in quantum field theory is ultraviolet divergent. For various physical reasons,  it is expected that the corresponding quantity  in a more complete theory including gravity should be finite. Finiteness of entanglement entropy is  of fundamental importance with implications for the black hole information paradox or holography. Since the modular integrals in string theory are naturally rendered finite in the ultraviolet, the partition function above is automatically ultraviolet finite and one may hope for a systematic computation of finite entanglement entropy order by order in perturbation theory. In the absence of a more direct way for defining and computing entanglement entropy, the orbifold method offers a promising route to address this important problem. 

One of the hurdles in implementing this idea is the fact that the spectrum of the $\mathbb{Z}_N$ orbifolds contains many tachyons in the twisted sectors and consequently the modular integral, even though finite in the ultraviolet,  is divergent in the infrared. This divergence of the integral in itself need not be a cause for despair  since the regime $N>1$ is not the physical regime in which one expects to be able to define R\'enyi or von Neumann entropy. It would suffice if the integrand, which is well defined over the modular domain, could be analytically continued to the physical region such that the resulting integral is finite in the physical domain $0< N \leq 1$. 

Such an approach was adopted by Witten \cite{Witten:2018xfj} to analyze the simpler problem of open strings on the Rindler horizon corresponding to the entanglement entropy on the D-brane worldvolume. Open strings offer two simplifications compared to closed strings. First, there are no twisted sectors for open strings. As a result, the one-loop annulus partition function for open strings has only $N$ sectors corresponding to the $N$ twines compared to the $N^2$ sectors corresponding to $N$ twists and $N$ twines for closed strings. Second, since the boundary conditions at the ends of open string relate left-movers to right-movers, one has to deal with only a single set of oscillators. 

Using a generalization of the Sommerfeld-Watson method, Witten \cite{Witten:2018xfj}  was able to obtain an explicit analytic continuation of the integrand for the annulus partition function. Its uniqueness is guaranteed by the Carlson theorem \cite{Carlson1914, Boas:1954}.  
The annulus  can be viewed  as the conformally equivalent cylinder corresponding to tree-level closed string exchange between D-branes allowing one to access the closed string sector.  Encouragingly, one finds that the closed string tachyonic terms which are present for $N >1$ are no longer present in the physical region $0 < N \leq 1$. 

The corresponding computation for closed strings is considerably more difficult. Even though the analytic continuation of the entire partition function is currently not available, it was observed in \cite{Dabholkar:2023ows} that one can  argue for the finiteness of the entanglement entropy by considering the analytic continuation of the divergent tachyonic terms. This argument relies on the fact  the tachyons responsible for the infrared divergence have a very specific structure in string theory. It was shown that all tachyonic terms from twisted sectors of the closed string in the modular integrand can be resummed to a form which can be analytically continued to $0 < N \leq 1$ and the resulting function is finite in the infrared.  After subtracting the tachyonic contribution,  the remaining part of the integrand is manifestly finite for odd positive values of $N$ and can be integrated and extrapolated to the physical domain.

Toroidal compactification of the ten-dimensional theory can give models in lower dimensions with finite and calculable entanglement entropy which will depend on the moduli of the compactification. 
It is natural to ask whether this method can be extended to   generic compactifications with less supersymmetry. The resummation of tachyonic terms relies on a number of specific features of the tachyonic spectrum and it is not immediately obvious that these features will persist in a generic compactification. In particular, in a generic orbifold compactification,  one expects additional tachyons from doubly twisted sectors which can have a very different structure and need not admit a resummation and finite analytic continuation.

With this motivation we consider a number of illustrative examples of Calabi-Yau  compactifications  to four dimensions and to six dimensions in orbifold limits. 
In \S\ref{sec-review} we review  the results of \cite{Dabholkar:2023ows} and  the resummation of the tachyons in ten dimensions.  In \S\ref{sec-cy}, we consider the compactification to four dimensions on a $T^6/\mathbb{Z}_3$ orbifold and find that there are no additional tachyons in the spectrum. In \S\ref{sec-k3} we study the compactification to six dimensions on $T^4/\mathbb{Z}_3$ and find that now there are additional tachyons  in the doubly-twisted sectors but which nevertheless can be resummed as in ten dimensions to obtain finite entropy. Some details of resummation are quite different from the ten-dimensional  case  suggesting that the finiteness of entropy is a generic feature in string theory. In \S\ref{sec-other} we summarize the results for other orbifold limits of $\mathbf{K3}$ and briefly discuss the Neveu-Schwarz-Ramond formalism and general CY compactifications.

\section{Analytic Continuation of Tachyons}
\label{sec-review}

In this section,  we review earlier work \cite{Dabholkar:2023ows} where it was argued that the entanglement entropy in the ten-dimensional Type-II superstring theory is finite.   The starting point is the one-loop partition function for the $\mathbb{Z}_N$ orbifold.  For a string background with $d$ non-compact directions, it is given by
\be
Z^{(1)} (N) = A_H^{(d-2)}  \int\limits_{\mathcal{D}} \frac{\dd[2]{\tau}}{\tau_2^2} \mathcal{F} (\tau, N)  , \label{modint}
\ee
where $A_H^{(d-2)}$ is the regularized  area in string units of the transverse $(d-2)$-dimensional  horizon and $\mathcal{D}$ is the standard fundamental domain in the upper-half-$\tau$ plane with $\tau \equiv \tau_1 + \mi \tau_2$. {To be more precise,  let $V^{(d-2)}$ be the volume of the transverse $(d-2)$-dimensional flat manifold.  The area $A_H^{(d-2)}$ mentioned above is then defined as $V^{(d-2)}/(2\pi l_s)^{d-2}$,  where $l_s$ is the string length. We work in the limit of very large $V^{(d-2)}$ so that \eqref{modint} is the leading term in an expansion in large $A_H^{(d-2)}$.  Furthermore,  this represents the stringy one-loop correction to the leading order answer which goes as $1/g_s^2$, $g_s$ being the string coupling.  This particular term in the genus expansion is therefore independent of $g_s$. }

For the ten-dimensional theory, the integrand $\mathcal{F} (\tau, N)$ is given in terms of the Jacobi theta and Dedekind eta functions 
\cite{Dabholkar:2023ows,  Dabholkar:2022mxo} by \be 
\mathcal{F} (\tau, N) = \frac{1}{N \tau_2^3}  \sum_{ \substack{ k, \ell \in \mathbb{Z}_N \\ (k, \ell) \neq (0,0) }  }   \left| \frac{\vartheta^4( \frac{k\tau + \ell}{N}  | \tau  )}{\eta^9 (\tau) \vartheta( \frac{2k\tau + 2\ell}{N}  | \tau  )} \right|^2,  \label{thet10d}
\ee
following the same conventions as in \cite{Dabholkar:2023ows,  Dabholkar:2022mxo}.   The $(k, \ell) = (0,0)$ term is absent because of the supersymmetry of the parent theory. The results described below follow easily from the asymptotics of the $\vartheta$ and $\eta$ functions, but we follow here instead the Hamiltonian approach to gain a better understanding in terms of physical states.

For $N=1$, the spacetime theory is maximally supersymmetric and the function \eqref{modint} vanishes identically.  For odd $N>1$,  supersymmetry is broken completely and there are localized tachyons in the twisted sectors. The integral is divergent in the infrared because of contributions of these tachyonic states to $\mathcal{F} (\tau, N)$  which grow exponentially in $\tau_2$. We now recall the structure of the tachyonic spectrum.   

  It is convenient to use the Green-Schwarz formalism in the light-cone gauge.  We number and pair the  eight transverse directions as
\be 
(01)(23)(45)(67) \label{pair4}
\ee
and take the Euclidean Rindler plane $\mathbb{R}^2$ to be along the $(01)$ directions. The $\mathbb{Z}_N$ orbifold group is generated by the element $g = \exp( 4 \pi \mi J_{01}/ N   )$, where $J_{01}$ is the generator of rotation in the $(01)$ plane.  Each pair of real bosons can be combined into a complex boson giving four complex bosons $X$ and   $\{Y^{i}\} \,  (i=1,2,3)$ where  $X$ corresponds to the $(01)$ directions and  $\{ Y^i \}$ correspond to the transverse $(234567)$ directions. In addition, there are four pairs of complex fermions $S^{a}$ and   $\tilde{S}^a$  ($a=1,2,3,4$),  corresponding to right- and left-movers, respectively. 

  Let $\Phi (\sigma^1, \sigma^2)$ be a general world-sheet field with the {twisted} boundary condition,
\be 
\Phi ( \sigma^1 +2\pi  ,  \sigma^2 ) = e^{2 \pi \mi \theta } \Phi (\sigma^1 , \sigma^2), \label{phibc}
\ee
 with $\sigma^1$,  $\sigma^2$ the spatial and temporal directions of the Euclidean world-sheet.
The twist $\theta$  can be taken to  lie in the range $[0,1)$ by an appropriate shift since the twisted boundary condition depends only on $\theta\, \, \mathrm{mod} \, 1$. 
For a (chiral) complex boson or a fermion twisted by $\theta$ with $ 0 \leq \theta < 1$,  the ground state energies are given respectively by \cite{Dixon:1985jw, Dixon:1986jc},
\be 
\begin{aligned}
\e_{\mathrm{B}} &=  - \frac{1}{12} + \frac{1}{2} \theta (1-\theta),  \\
\e_{\mathrm{F}} &=  + \frac{1}{12} - \frac{1}{2} \theta (1-\theta), \label{gsform}
\end{aligned}
\ee
for the right movers and similarly for the left movers.

The twisted sectors of the orbifold are labeled by the integer $k$,  $0 \leq k < N-1$ where $k=0$ is the untwisted sector.  The twists $\theta$ for the world-sheet fields $(X,  Y^i,  S^a)$ in the $k$-twisted sector are given by $(2k/N,  \,  0,  \,  k/N)$.  Therefore,  using the formul\ae\ \eqref{gsform} and adding contributions of all fields both for left and right movers, one finds that the total energy of the ground state is given by
\be 
\e = \begin{cases} 
- \frac{2k}{N} ,  &\quad \qty(1 \leq k \leq \frac{N-1}{2}), \\
- \frac{2(N-k)}{N} ,  &\quad \qty(\frac{N+1}{2} \leq k \leq N-1).  \label{rtch1} \, 
\end{cases}
\ee

Since $\e$ is negative,  the ground state corresponds to a tachyon  in the spacetime interpretation. We refer to it as the \emph{leading tachyon} since it has the most negative mass-squared in  the $k$-twisted sector.  In addition, there are many \emph{sub-leading tachyons} obtained by acting with fractionally-moded oscillators on the ground state. Requiring level-matching and $\mathbb{Z}_N$-invariance, only the fractionally-moded $X$ oscillators give rise to sub-leading tachyons obtained by repeated application of  
$\alpha_{-1 + \frac{2k}{N} }^X  \widetilde{\overline{\alpha}}^X_{-1 + \frac{2k}{N} }$ with an additional energy $2(1-2k/N)$, for $1 \leq k \leq (N-1)/2$ (and similar oscillators in the range $(N + 1)/ 2 \leq k \leq N -1$).   It is thus clear that all tachyonic states in a  given sector have unit degeneracy.

Noting a $k \leftrightarrow N- k$ symmetry of the tachyonic  spectrum (see,  e.g.,  \eqref{rtch1}),  the total tachyonic contribution to $\cF_0(\tau_2, N)$ is thus given by
\be 
\mathcal{F}^T_0 (\tau_2 ,  N) = \frac{2}{\tau_2^3} \sum_{ k =1}^{ \frac{N-1}{2} } \exp( \frac{4\pi \tau_2 k}{N}  )\sum_{n =0}^{n_k} \exp( - 4 \pi \tau_2 n \pqty{ 1- \frac{2k}{N} }   ),\label{ft0}
\ee
where $n_k$ is a non-negative integer such that the above sum includes only tachyons and no massless or massive states.  The presence of this $k$-dependent integer makes it difficult to resum it in a form suitable for analytic continuation in $N$.  To circumvent this difficulty,  we consider a modified sum $\widetilde{\mathcal{F}}^T_0 (\tau_2, N)$ by taking the limit $n_k \to \infty$,  which corresponds to adding contributions from some massless and infinitely many massive states.  One can now change the order of the $k$ and $n$ sums.  The $k$-sum for each $n$ is a finite geometric series which can be readily summed to obtain
\be
\widetilde{\mathcal{F}}^T_{0,1} (\tau_2 ,  N) = - \frac{2}{\tau_2^3} \sum_{n=0}^\infty e^{ - 4 \pi n \tau_2} \frac{1 - \exp(\frac{N-1}{N} (2 n+1) 2 \pi \tau _2 )  }{1 - \exp(-\frac{1}{N} (2 n+1) 4 \pi \tau _2 ) }. \label{ft02}
\ee
The sum above admits a natural analytic continuation to the physical region  $0 < N \leq 1$.  We have added an additional suffix ``$1$'' here for later convenience. Remarkably,  although the function $\widetilde{\mathcal{F}} ^T_{0,1} (\tau_2 ,  N)$ diverges as $\tau_2\to \infty$ when $N>1$,  it remains finite in  the physical region $0 < N \leq 1$.

To argue for the finiteness of the full partition function,  we separate the tachyonic and non-tachyonic parts in $\mathcal{F} (\tau, N)$ as
\be 
\mathcal{F} (\tau, N) = \widetilde{\mathcal{F}}^T_{0,1} (\tau_2 ,  N) +   \widetilde{\mathcal{F}}^R (\tau ,  N), \label{fsep}
\ee
where the remainder term $ \widetilde{\mathcal{F}}^R (\tau ,  N)$ is tachyon-free for $N=1,3,5,\cdots$ and hence the modular integral \eqref{modint} involving only $\widetilde{\mathcal{F}}^R (\tau_2 ,  N)$ is finite for these values.  One can extrapolate the modular integral to values of $0 < N \leq 1$, which is expected to be finite.   The modular integral of $\widetilde{\mathcal{F}}^T_{0,1} (\tau_2 ,  N)$ is manifestly finite in this region.  Therefore,  the full modular integral \eqref{modint} is finite in $0< N \leq 1$.  { As mentioned in \cite{Dabholkar:2023ows},  there is a pathological possibility that new tachyonic divergences might appear in the analytic continuation of the finite remainder in the physical region.  Given that the finite part corresponds to only massless and massive fields,  realization of  such a possibility seems quite unlikely.  We hope to return to this question in a future work.}

\section{\boldmath Compactification on ${T^6}/\mathbb{Z}_3$}\label{sec-cy}

The resummation and analytic continuation of the tachyonic terms summarized above  depends crucially on some non-trivial and specific features of the ten-dimensional Type-II superstring theory. It is  natural to ask if these features are shared by a more general string compactification so that the analytic continuation continues to hold.
To examine this question concretely,  we first consider 
the compactification on the orbifold $T^6/\mathbb{Z}_3$ to illustrate the main points. 

We continue to work with the Green-Schwarz formalism in the light-cone gauge and take  the six directions $(23)(45)(67)$ transverse to the Rindler plane to be the toroidal directions. 
 When the  $T^6$ is of the special form $T^6 = T^2 \times T^2 \times T^2$ with each $T^2$ as the maximal torus of $SU(3)$, it admits a $\mathbb{Z}_3$ symmetry generated by 
 \be
T = \exp( \frac{2\pi \mi}{3} (J_{23} + J_{45} - 2J_{67})  ).  \label{cy3gen}
\ee
which has unit determinant and is an element of $SU(3) \in    \mathrm{Spin}(6)$. Hence, the orbifold $T^6/\mathbb{Z}_3$ can be viewed as a limit of a Calabi-Yau threefold with $SU(3)$ holonomy, and the resulting four-dimensional theory has ${\cal N} = 2$ supersymmetry. 

We have four complex bosons $X$,  $Y^i$ ($i=1,2,3$)  and  four complex fermions $S^{++++}$,  $S^{++--}$,  $S^{+-+-}$ and $S^{+--+}$ where the superscripts denote the (half-integral) charges along the four planes.  
The two generators $g$ and $T$ generate the orbifold group $\mathbb{Z}_N \times \mathbb{Z}_3$. We label the  $\mathbb{Z}_N$ twists and twines with $(k, \ell)$ each taking values in $\{0, 1,\cdots N-1\}$ and the $\mathbb{Z}_3$ twists and twines  with $(r, s)$ taking values in $\{0, 1,2\}$. 
The boundary conditions in sectors twisted by $k$ and $r$ are
\be
\begin{aligned}
X (\sigma^1 + 2\pi,  \sigma^2) &= \exp(2 \pi \mi \frac{2 k}{N}) X(\sigma^1,  \sigma^2),   \\
Y^1 (\sigma^1 + 2\pi,  \sigma^2  ) &= \exp(2\pi \mi  \frac{r}{3}) Y^1 (\sigma^1,  \sigma^2)  ,   \\
Y^2 (\sigma^1 + 2\pi,  \sigma^2) &= \exp(2\pi \mi  \frac{r}{3}) Y^2 (\sigma^1,  \sigma^2) , \\
Y^3 (\sigma^1 + 2\pi,  \sigma^2) &= \exp(- 2\pi \mi \frac{ 2r}{3}) Y^3 (\sigma^1,  \sigma^2) \label{cy3bos}
\end{aligned}
\ee
for bosons and \be
\begin{aligned}
S^{++++} (\sigma^1 +2\pi,  \sigma^2) &= \exp(2 \pi \mi \frac{k}{N}) S^{++++} (\sigma^1,  \sigma^2) , \\
S^{++--} (\sigma^1 +2\pi,  \sigma^2) &= \exp[2 \pi \mi \pqty{ \frac{k}{N}  + \frac{r}{3} } ] S^{++--} (\sigma^1,  \sigma^2),  \\
S^{+-+-} (\sigma^1 +2\pi,  \sigma^2) &= \exp[2 \pi \mi  \pqty{ \frac{k}{N} + \frac{r}{3} }]  S^{+-+-} (\sigma^1,  \sigma^2),  \\
S^{+--+}(\sigma^1 +2\pi,  \sigma^2) &= \exp[2 \pi \mi  \pqty{ \frac{k}{N} - \frac{2r}{3} }]  S^{+--+} (\sigma^1,  \sigma^2) \label{cy3fer}
\end{aligned}
\ee
for fermions. The twines $\ell$ and $s$ are similarly implemented by twisted boundary conditions along the $\sigma^2$-direction. 
One can then write down the partition function by inspection \cite{Alvarez-Gaume:1986rcs}:
\be
\begin{aligned}
\mathcal{F} (\tau,  N) &= \frac{1}{3N} \sum_{ \substack{ k, \ell \in \mathbb{Z}_N \\ (k, \ell) \neq (0,0) }  }   \left| \frac{\vartheta^4( \frac{k\tau + \ell}{N}  | \tau  )}{\eta^9 (\tau) \vartheta( \frac{2k\tau + 2\ell}{N}  | \tau  )} \right|^2 \sum_{ ( \mathbf{p}_R,  \mathbf{p}_L) \in \Gamma_{6,6}  } q^{  \mathbf{p}_R^2/2 } \overline{q}^{  \mathbf{p}_L^2/2 }  \\
&\quad + \frac{1}{3N} \sum_{ \substack{ k, \ell \in \mathbb{Z}_N \\ r, s \in \mathbb{Z}_3 \\(r,s) \neq (0,0) \\  (k, \ell) \neq (0,0)   } } \chi(r,s)   \left|  \frac{\vartheta( \frac{k\tau + \ell}{N}  | \tau  )  \vartheta^2 ( \frac{k\tau + \ell}{N} + \frac{r \tau + s}{3}  | \tau  ) \vartheta ( \frac{k\tau + \ell}{N} - \frac{2r \tau + 2s}{3}  | \tau  ) }{\vartheta( \frac{2k\tau + 2\ell}{N}  | \tau  ) \vartheta^2 ( \frac{r\tau + s}{3}  | \tau  ) \vartheta ( \frac{2r\tau + 2s}{3}  | \tau  ) } \right|^2 \label{pfcy3},
 \end{aligned}
 \ee
 where $\Gamma_{6,6}$ is the Narain lattice \cite{Narain:1985jj, Narain:1986am} for the torus $T^6$ and $q \equiv \exp(2\pi \mi \tau)$.  The term in the first line above corresponds to the $r= 0$ and  $s=0$ sector of the  $r,s$ sum, which is the only term that depends on the Narain lattice.  
 From the knowledge of $\chi(0,s)$ in the untwisted sector, one can determine $\chi(r,s)$ by modular invariance.  For the $\mathbb{Z}_3$ orbifold,  $\chi(r,s) = 27$ for all values of $r$ and $s$, corresponding to the number of $\mathbb{Z}_3$ fixed points on $T^6$. 
 
 From the large $\tau_2$ asymptotics,  one can easily identify the exponentially growing tachyonic terms in this partition sum.
 Below we analyze these tachyonic terms from the operator formalism to gain a better understanding of the structure.  
 
\subsection[The $r=0$ Untwisted Sector]{\boldmath The $r=0$ Untwisted Sector}\label{subsec-cy3r0}

The spectrum and the partition function in this sector is moduli-dependent because of the Narain lattice sum for the $T^6$ factor in the first line of \eqref{pfcy3}. The $T^6$ partition function is multiplied by exactly the same combination of $\vartheta$ and $\eta$ functions appearing in the ten-dimensional case \eqref{thet10d}.  Therefore,  this combination gives rise to exactly the same leading and sub-leading tachyons except that they now appear as a Kaluza-Klein tower of four-dimensional tachyons.

 Since the $T^6$  partition function is $N$-independent, it factorizes, and the $N$-dependent part of the first term is exactly the same as \eqref{thet10d}. As a result,  the tachyonic resummation and analytic continuation works essentially the same way as in \S\ref{sec-review}. 
This is to be expected since  the  Kaluza-Klein tower of tachyons is simply the  four-dimensional repackaging of the ten-dimensional states.

It is easy to check that in the $\mathbb{Z}_3$-projected sectors,  $s \neq 0$,  
one obtains the same tachyons as in \S\ref{sec-review}, with an overall constant factor due to the $s$-sum. The ground states in the remaining sectors  have to be analyzed carefully sector by sector.

\subsection[The $r=1$ Twisted Sector]{\boldmath The $r=1$ Twisted Sector}\label{subsec-cy3r1}

There are three sub-cases as below.

\begin{itemize}
\item \underline{\it Subcase 1}:  $0< k/N < \frac{1}{2}$

In this case,  the bosonic twists \eqref{cy3bos} in the standard interval are given by $2k/N$,  $1/3$, $1/3$  and $1/3$ $(= -2/3 +1)$ respectively. The total bosonic ground state energy using \eqref{gsform} is given by
\be 
\begin{aligned}
\e_{\mathrm{B}} &= - \frac{4}{12} + \frac{1}{2} \frac{2k}{N} \pqty{ 1 - \frac{2k}{N}} + 3 \times \frac{1}{2} \times \frac{1}{3} \times \pqty{ 1 - \frac{1}{3} } \\
&= \frac{k}{N} \pqty{ 1 - \frac{2k}{N}} .
\end{aligned} \label{eb3k1}
\ee
The fermions,  on the other hand,  are twisted by $k/N$,  $k/N + 1/3$,  $k/N + 1/3$ and $k/N + 1/3$ $(= k/N - 2/3 +1)$ respectively,  with ground state energy
\be 
\begin{aligned}
\e_{\mathrm{F}} &= \frac{4}{12} - \frac{1}{2} \frac{k}{N} \pqty{ 1 - \frac{k}{N}} - 3 \times \frac{1}{2} \times \pqty{ \frac{k}{N} + \frac{1}{3} } \pqty{1 - \frac{k}{N} - \frac{1}{3} } \\
&= -\frac{k}{N} \pqty{ 1 - \frac{2k}{N}} .
\end{aligned} \label{ef3k1}
\ee

Thus,  the bosonic and fermionic energies cancel each other for this range of values of $k$ and the total ground state energy vanishes,
\be 
\e = 0 \, ,  \label{cy3e13}
\ee
corresponding to a massless state in spacetime. 

\item \underline{\it Subcase 2}:  $\frac{1}{2}< k/N < \frac{2}{3}$

The bosonic twists \eqref{cy3bos} in the standard domain are $2k/N -1$,  $1/3$, $1/3$ and $1/3$ respectively leading to, 
\be 
\e_{\mathrm{B}} = -\frac{2 k^2}{N^2}+\frac{3 k}{N}-1,  \label{ebcy3ck2}
\ee
while the fermionic twists remain the same as in the previous subcase with  the same energy \eqref{ef3k1}.  The total ground state energy including both the right- and left-movers is given by
\be 
\e =2 \pqty{ \frac{2k}{N} - 1 },
\ee
corresponding to a non-tachyonic massive state in spacetime. 

\item \underline{\it Subcase 3}:  $\frac{2}{3} \leq k/N < 1$

The bosonic twists are unchanged with respect to the previous subcase and so is the bosonic ground state energy \eqref{ebcy3ck2}.  The fermionic twists, on the other hand,  are given by $k/N$,  $k/N -2/3$,  $k/N - 2/3$ and $k/N - 2/3$ respectively, giving
\be 
\e_{\mathrm{F}} = \frac{2 k^2}{N^2}-\frac{4 k}{N}+2.  \label{efcy3k3}
\ee
The total energy is thus
\be 
\e = 2 \pqty{1 - \frac{k}{N} }, \label{etotcy33}
\ee
again corresponding to a  massive state.
\end{itemize}

\subsection[The $r=2$ Twisted Sector]{\boldmath The $r=2$ Twisted Sector}\label{subsec-cy3r2}

In this sector there are three different ranges of $k$ listed below.

\begin{itemize}

\item \underline{\it Subcase 1}:  $0< k/N \leq \frac{1}{3}$ 

The bosonic twists in the domain $[0,1)$ are $2k/N$, $2/3$, $2/3$ and $2/3$ respectively, leading to the same value of energy as \eqref{eb3k1}.  The standard fermionic twists are $k/N$,  $k/N + 2/3$,  $k/N + 2/3$ and $k/N + 2/3$ respectively, leading to,
\be
\e_{\mathrm{F}} =  \frac{2 k^2}{N^2}.
\ee
Therefore,  we find the total energy
\be 
\e = \frac{2k}{N},
\ee
corresponding to a non-tachyonic massive state.

\item \underline{\it Subcase 2}:  $\frac{1}{3}< k/N < \frac{1}{2}$ 

The bosonic twists are unchanged with respect to the previous subcase and the energy continues to be given by \eqref{eb3k1}.  The fermionic twists are now $k/N$,  $k/N -1/3$,  $k/N - 1/3$ and $k/N - 1/3$ respectively, leading to,
\be 
\e_{\mathrm{F}} = \frac{2 k^2}{N^2}-\frac{3 k}{N}+1 .  \label{efcy3r23}
\ee
Thus,  the total energy is again positive
\be 
\e = 2 \pqty{ 1 - \frac{2k}{N} },
\ee
corresponding to a massive state.

\item \underline{\it Subcase 3}:  $\frac{1}{2}< k/N <1$

In this case, the bosonic energy is given by the expression \eqref{ebcy3ck2} and the fermionic energy continues to be given by \eqref{efcy3r23} so the total energy vanishes and the ground state corresponds to a massless state.
\end{itemize}
These results could be anticipated by noting that under the transformation $k \to N - k$ and $r \to 3 - r$,  the worldsheet fields are transformed into their complex conjugates.

To summarize,   for $r=1$,
\be
\e = \begin{cases}
0 ,  &\quad \qty( 0 \leq k < \frac{1}{2} ), \\
2 \pqty{ \frac{2k}{N} - 1} ,  &\quad \qty( \frac12 < k \leq \frac{2}{3} ), \\
2 \pqty{1-\frac{k}{N}} ,  &\quad \qty( \frac{2}{3}< k <1 )\, ;  \label{efincy31}
\end{cases}
\ee
while for $r=2$,
\be 
\e = \begin{cases}
\frac{2k}{N} ,  &\quad \qty( 0 \leq \frac{k}{N} \leq \frac{1}{3} ), \\
2\pqty{1- \frac{2k}{N}} ,  &\quad \qty( \frac{1}{3} < \frac{k}{N} < \frac{1}{2} ), \\
0 ,  &\quad \qty( \frac{1}{2}< \frac{k}{N} <1 ).
\end{cases} \label{efincy32}
\ee
Thus, this orbifold is completely free of any new tachyons apart from the ones we had already encountered in  \S\ref{sec-review}.  It is worth pointing out that the $k \leftrightarrow N-k$ symmetry continues to exist for the ground state spectrum but with a non-trivial rearrangement between the $r=2$ and $r=3$ sectors.

The tachyonic contribution to the partition function is given by,
\be 
\widetilde{\mathcal{F}}^T_0 (\tau_2 ,  N) = \tau_2^3 \widetilde{\mathcal{F}}^T_{0,1} (\tau_2 ,  N) \bqty{ \frac{1}{3} Z(\Gamma_{6,6})  +  \frac{2}{3} },  \label{cy3tach}
\ee 
where $Z(\Gamma_{6,6}) = \sum_{\Gamma_{6,6}  } q^{  \mathbf{p}_R^2/2 } \overline{q}^{  \mathbf{p}_L^2/2 } $ is the Narain  partition function that appears in \eqref{pfcy3} and $\widetilde{\mathcal{F}}^T_{0,1} (\tau_2 ,  N)$ is the same function we had encountered in \eqref{ft02}.

\section{\boldmath Compactification on $T^4/\mathbb{Z}_3$ \label{sec-k3}}

In the example considered in the previous section, we found no additional tachyons from the doubly-twisted sectors . This feature does not persist for generic string compactifications. 
To illustrate this possibility, we consider orbifold compactifications to six dimensions on  $T^4/ \mathbb{Z}_M$ for $M=2,3,4,6$. We take the $(45)$ and $(67)$ directions in \eqref{pair4} to be along the $T^4$. 
 When the  $T^4$ is of the special form $T^4 = T^2 \times T^2 $ with each $T^2$  admitting a $\mathbb{Z}_M$ symmetry, the orbiold  generator can be taken to be 
  \be 
T = \exp( \frac{2\pi \mi}{M} (J_{45} - J_{67})  ),\label{gent4}
\ee
with $M=2, 3, 4, 6$ which has unit determinant and is an element of $SU(2) \in   \mathrm{Spin}(4)$. The resulting orbifold $T^4/\mathbb{Z}_M$ has $SU(2)$ holonomy which can be regarded as a limit of a $\mathbf{K3}$ and the resulting six-dimensional theory has ${\cal N} = 2$ supersymmetry.

The boundary conditions  in  the $k$ and $r$ twisted sectors  are given by
\be
\begin{aligned}
X (\sigma^1  +2 \pi,  \sigma^2) &= \exp(2 \pi \mi \frac{2 k}{N}) X(\sigma^1,  \sigma^2),  \quad  \\
Y^1 (\sigma^1 +2 \pi,  \sigma^2 ) &= Y^1 (\sigma^1,  \sigma^2) , \\
Y^2 (\sigma^1  + 2\pi,  \sigma^2) &= \exp(2\pi \mi  \frac{r}{M}) Y^2 (\sigma^1,  \sigma^2) ,  \\
Y^3 (\sigma^1  + 2\pi,  \sigma^2) &= \exp(- 2\pi \mi \frac{ r}{M}) Y^3 (\sigma^1,  \sigma^2) \label{zmbos1}
\end{aligned}
\ee
for the bosons, and
\be
\begin{aligned}
S^{++++} (\sigma^1 +2\pi,  \sigma^2) &= \exp(2 \pi \mi \frac{k}{N}) S^{++++} (\sigma^1,  \sigma^2), \\
S^{++--}  (\sigma^1 +2\pi,  \sigma^2) &= \exp(2 \pi \mi \frac{k}{N}) S^{++--} (\sigma^1,  \sigma^2) , \\
S^{+-+-} (\sigma^1 +2\pi,  \sigma^2) &= \exp[2 \pi \mi  \pqty{ \frac{k}{N} + \frac{r}{M} }]  S^{+-+-} (\sigma^1,  \sigma^2) ,\\
S^{+--+}  (\sigma^1 +2\pi,  \sigma^2) &= \exp[2 \pi \mi  \pqty{ \frac{k}{N} - \frac{r}{M} }]  S^{+--+}(\sigma^1,  \sigma^2) \label{zmfer1}
\end{aligned}
\ee
for the fermions. 
As before, the partition function can be written down by inspection:
\be 
\begin{aligned}
\mathcal{F} (\tau,  N, M) &=  \frac{1}{MN}  \sum_{ \substack{ k, \ell \in \mathbb{Z}_N \\ (k, \ell) \neq (0,0) }  }  \frac{1}{\tau_2} \left| \frac{\vartheta^4( \frac{k\tau + \ell}{N}  | \tau  )}{\eta^9 (\tau) \vartheta( \frac{2k\tau + 2\ell}{N}  | \tau  )} \right|^2 \sum_{ ( \mathbf{p}_R,  \mathbf{p}_L) \in \Gamma_{4,4}  } q^{  \mathbf{p}_R^2/2 } \overline{q}^{  \mathbf{p}_L^2/2 }  \\
&\quad + \frac{1}{MN} \sum_{ \substack{ k, \ell \in \mathbb{Z}_N \\ r, s \in \mathbb{Z}_M \\ (r,s) \neq (0,0) \\ (k, \ell) \neq (0,0) } } \frac{ \chi(r,s)  }{\tau_2} \left|  \frac{\vartheta^2 \pqty{ \frac{k\tau + \ell}{N}  | \tau }  \vartheta \pqty{ \frac{k\tau + \ell}{N} + \frac{r\tau + s}{M}  | \tau } \vartheta \pqty{ \frac{k\tau + \ell}{N} - \frac{r\tau + s}{M}  | \tau } }{\eta^3 (\tau) \vartheta \pqty{ \frac{2k\tau + 2\ell}{N}  | \tau } \vartheta^2 \pqty{ \frac{r\tau + s}{M}   | \tau } } \right|^2 .  \label{pfk3m}
\end{aligned}
\ee
In the untwisted sector,  $\chi ( 0 , s) = 16 \sin^4 (\pi s/ M)$.  Modular invariance fixes all other values of $\chi(r,s)$. 

We now describe the $M = 3$ case in some detail and summarize the other cases in the next section.  
The $\mathbb{Z}_3$-untwisted sector,  $r=0$, can be treated similarly as before.   In this case, $\chi (r, s ) = 9$ for all $r$ and $s$.
For the $\mathbb{Z}_3$-twisted sectors $r =1, 2$,  the ground state energies are: 
\be
\e = \begin{cases}
0,  &\quad \qty( 0 \leq  \frac{k}{N} < \frac{1}{3} ),  \\
2 \pqty{ \frac{1}{3} - \frac{k}{N}},  &\quad \qty( \frac{1}{3} \leq  \frac{k}{N} < \frac{1}{2} ),  \\
2\pqty{ \frac{1}{3} - \frac{N-k}{N} },  &\quad \qty( \frac{1}{2} < \frac{k}{N} \leq \frac{2}{3} ),  \\
0,  &\quad \qty( \frac{2}{3} < \frac{k}{N} < 1).
\end{cases} \label{letaz3}
\ee
 The ground state energies in the range $1/3 < k/N < 2/3$ are negative and therefore these states correspond to spacetime tachyons.  There is a $k \leftrightarrow N - k$ symmetry for each value of $r$ and an $r \leftrightarrow M -r$ symmetry. 

As before, one obtains sub-leading tachyons by an application of the fractionally moded $X$ oscillators $\alpha_{-1 + \frac{2k}{N} }^X  \widetilde{\overline{\alpha}}^X_{-1 + \frac{2k}{N} }$  (in the range $1 \leq k \leq (N-1)/2$).  Further analysis shows that there are no other oscillators which can produce sub-leading tachyons after imposing level-matching and $\mathbb{Z}_3$-invariance. 
 The tachyonic contribution for each of the $r=1$ and $r=2$ sectors is given by 
 \be 
\mathcal{F}^T_{0,  3} (\tau_2,  N) = \frac{2}{\tau_2} \sum_{ k = \lfloor \frac{N}{3} \rfloor + 1 }^{ \frac{N-1}{2} } \exp(- 4 \pi \tau_2 \pqty{ \frac{1}{3} - \frac{k}{N} }  ) \sum_{n =0}^{n_k} \exp( - 4 \pi \tau_2 n \pqty{ 1- \frac{2k}{N} }    ).  \label{sumz3}
\ee
Taking $n_k \to \infty$ and interchanging the sums,  we obtain a finite geometric series in the inner sum.  
However, replacing the floor function $\lfloor \frac{N}{3} \rfloor$ by $0$ adds only nontachyonic terms. Hence, one can take the sum to go from $k=1$ to $k = (N-1)/2$ and obtain
\be 
\widetilde{\mathcal{F}}^T_{0,  3} (\tau_2,  N) = - \frac{2}{\tau_2} \sum_{n=0}^\infty e^{- \pqty{2n+ \frac{2}{3}} 2 \pi \tau_2 }  \frac{1 - \exp(\frac{N-1}{N} (2 n+1) 2 \pi \tau _2 )  }{1 - \exp(-\frac{1}{N} (2 n+1) 4 \pi \tau _2 ) } \label{fo3k1} \,.
\ee
This expression makes it clear that,  as long as we include \emph{all the tachyons},  the finiteness in the physical regime is determined by the upper limit of the $k$-sum in \eqref{sumz3}.  The finiteness of $\widetilde{\mathcal{F}}^T_{0,  3} (\tau_2,  N) $ is actually valid in the greater domain $0 < N \leq 3$.
The entire tachyonic contribution can then be summarized as
\be
\widetilde{\mathcal{F}}^T_{0} (\tau_2,  N,  3) =   \tau_2^2 \widetilde{\mathcal{F}}^T_{0,1} (\tau_2 ,  N) \bqty{ \frac{1}{3} Z (\Gamma_{4,4}) + \frac{2}{3} } + 18 \widetilde{\mathcal{F}}^T_{0,  3} (\tau_2,  N), \label{tacht4z3}
\ee
where $ \widetilde{\mathcal{F}}^T_{0,1} (\tau_2 ,  N) $ is the ten-dimensional tachyonic contribution as in \eqref{ft02}  and $Z (\Gamma_{4,4}) =  \sum_{\Gamma_{4,4}  } q^{  \mathbf{p}_R^2/2 } \overline{q}^{  \mathbf{p}_L^2/2 } $ is the Narain partition function on $T^4$. 
There is a factor of 18 in the second term because $\chi (r,s) =9 $ in both the $r=1$ and $r=2$ sectors.

The total partition function \eqref{pfk3m} can be written as before as \be 
\mathcal{F} (\tau, N,  3) = \widetilde{\mathcal{F}}^T_{0} (\tau_2,  N,  3) +   \widetilde{\mathcal{F}}^R (\tau,  N,  3).
\ee
We can similarly argue for the finiteness of the modular integral in $0 < N \leq 1$.

\section{Other Compactifications}\label{sec-other}
 
The boundary conditions for the bosons and fermions in the $T^4 / \mathbb{Z}_M$ orbifold are as in \eqref{zmbos1} and \eqref{zmfer1}, with the total partition function given by \eqref{pfk3m}. We summarize the results below and conclude this section with comments on  more general compactifications.

\subsection[$T^4 / \mathbb{Z}_2$]{\boldmath $T^4 / \mathbb{Z}_2$}\label{subsec-t4z2}

For $M = 2$,  in the doubly-twisted sector ($r = 1$), the total ground state energy vanishes and there are no additional tachyons in the doubly-twisted sector. 
The total tachyonic contribution for the $\mathbb{Z}_2$ orbifold is given by
\be 
\widetilde{\cal F}^T_{0} (\tau_2 , N,  2) =  \tau_2^2 \widetilde{\cal F}^T_{0, 1} (\tau_2 , N) \bqty{ \frac{1}{2} Z(\Gamma_{4,4}) + \frac{1}{2}} ,
\ee
where $ \widetilde{\cal F}^T_{0, 1} (\tau_2 , N)$ is defined by \eqref{ft02}.

\subsection[$T^4 / \mathbb{Z}_4$]{\boldmath $T^4 / \mathbb{Z}_4$}\label{subsec-t4z4}

The $r=2$ case for $M=4$ is similar to the $r=1$ case for $M=2$.  Thus, the ground state energy vanishes and there are no tachyons. 

For each of the $r=1$ and $r=3$ sectors,  we find that $\chi(r,s) = 4$ and we obtain the following expressions for the total energy of the ground state:
\be
\e = \begin{cases}
0,  &\quad \qty( 0 \leq  \frac{k}{N} < \frac{1}{3} ),  \\
2 \pqty{ \frac{1}{4} - \frac{k}{N}},  &\quad \qty( \frac{1}{4} \leq  \frac{k}{N} < \frac{1}{2} ),  \\
2\pqty{ \frac{1}{4} - \frac{N-k}{N} },  &\quad \qty( \frac{1}{2} < \frac{k}{N} \leq \frac{3}{4} ),  \\
0,  &\quad \qty( \frac{3}{4} < \frac{k}{N} < 1).
\end{cases}  \label{letaz4}
\ee
On these leading tachyons one can act with the oscillators $\alpha_{-1 + \frac{2k}{N} }^X  \widetilde{\overline{\alpha}}^X_{-1 + \frac{2k}{N} }$  (in the range $1 \leq k \leq (N-1)/2$; and similar oscillators in the range $(N+1)/2  \leq k \leq N-1$) to obtain the subleading tachyons. From each of this sector, one finds a sum similar to \eqref{sumz3}.
As in \eqref{fo3k1},  one can perform the sum starting from $k=1$ to obtain 
\be 
\widetilde{\mathcal{F}}^T_{0,  4} (\tau_2,  N) = - \frac{2}{\tau_2} \sum_{n=0}^\infty e^{ - (4n +1) \pi \tau_2 } \frac{1 - \exp(\frac{N-1}{N} (2 n+1) 2 \pi \tau _2 )  }{1 - \exp(-\frac{1}{N} (2 n+1) 4 \pi \tau _2 ) }. \label{fo4k1}
\ee
Under this analytic continuation,  the additional twisted tachyons disappear in the larger region $0 < N \leq 2$.   
The net contribution of all the tachyons is encoded in the quantity
\be
 \widetilde{\cal F}^T_{0} (\tau_2 , N, 4) =   \tau_2^2 \widetilde{\cal F}^T_{0, 1} (\tau_2 , N) \bqty{ \frac{1}{4} Z(\Gamma_{4,4}) + \frac{3}{4} } +8 \widetilde{\cal F}^T_{0, 4} (\tau_2 , N),\label{netconz4}
\ee
which we can add and subtract from \eqref{pfk3m} for $M=4$ to repeat the usual procedure of obtaining a finite total answer in $0 < N \leq 1$.

\subsection[$T^4 / \mathbb{Z}_6$]{\boldmath $T^4 / \mathbb{Z}_6$}\label{subsec-t4z6}

The $r=3$ sector is  similar to the $r=1$ sector of $M=2$ and contains no tachyons. On the other hand,  the $r = 2$ and $r=4$ sectors are  similar  to the $r=1,2$ sectors for $M=3$.  These sectors will have the tachyons as previously discussed.

What remain are the $r=1$ and $r=5$ sectors for $M=6$,  for which,
\be
\e = \begin{cases}
0,  &\quad \qty( 0 \leq  \frac{k}{N} < \frac{1}{3} ),  \\
2 \pqty{ \frac{1}{6} - \frac{k}{N}},  &\quad \qty( \frac{1}{6}<\frac{k}{N} < \frac{1}{2} ),  \\
2\pqty{ \frac{1}{6} - \frac{N-k}{N} },  &\quad \qty( \frac{1}{2} < \frac{k}{N} <\frac{5}{6} ),  \\
0,  &\quad \qty( \frac{5}{6} < \frac{k}{N} < 1). \label{letaz6}
\end{cases} 
\ee
The situation with the sub-leading tachyons is a bit more subtle in this case.  In the situations we had previously discussed,  there were no sub-leading tachyons when the $Y$-oscillators acted on the tachyonic vacuum.  In this case, by contrast,  we can act with the $Y$ oscillators: $\alpha^Y_{-1/6} \overline{\widetilde{\alpha}}^Y_{-1/6}$.  There are $4$ different distinct oscillators possible corresponding to the two $Y$-oscillators in each sector.  For the resulting state to remain tachyonic,  only one application of the pair of oscillators is allowed and the resulting spectrum is identical to \eqref{letaz3}.  One can then proceed with applications of the $X$-oscillators on top of it which yields a contribution proportional to \eqref{fo3k1}.  In this case,  $\chi(r,s) $ are not all equal. However,  the $\mathbb{Z}_3$ and $\mathbb{Z}_6$ tachyons can nevertheless be isolated for our purposes.

Acting on the ground states \eqref{letaz6} only with the $X$-oscillators and resumming the relevant terms as before,  one obtains 
\be 
\widetilde{\mathcal{F}}^T_{0,  6} (\tau_2,  N) = - \frac{2}{\tau_2} \sum_{n=0}^\infty e^{ - \pqty{ 2n + \frac{1}{3} } 2 \pi \tau_2} \frac{1 - \exp(\frac{N-1}{N} (2 n+1) 2 \pi \tau _2 )  }{1 - \exp(-\frac{1}{N} (2 n+1) 4 \pi \tau _2 ) } . \label{fo6k1}
\ee
This expression is finite as $\tau_2 \to \infty$,  for $0 < N \leq 3/2$.

The contribution from the tachyons from all possible sectors in analytically continued form is given by
\be 
 \widetilde{\cal F}^T_{0} (\tau_2 , N, 6) =  \tau_2^2 \widetilde{\cal F}^T_{0, 1} (\tau_2 , N) \bqty{ \frac{1}{6} Z(\Gamma_{4,4}) + \frac{5}{6} } +2 \widetilde{\cal F}^T_{0, 6} (\tau_2 , N) +18 \widetilde{\cal F}^T_{0, 3} (\tau_2 , N) .\label{netconz6}
\ee
The relative factors can be accounted for by considering the contributions of the $Y$-oscillators and $\chi(r,s)$.

\subsection{General Compactifications}\label{sec-conc}

We have so far used the Green-Schwarz formalism to simplify the calculations, which is well suited for orbifolds of free field theories.  This formalism is, however, not suitable when we discuss more general compactifications where the internal manifold is described by a sigma model or an abstract $\cN=2$ superconformal field theory (SCFT).   In such cases,  the Neveu-Schwarz-Ramond (NSR) formalism is more useful. An additional step in the NSR formalism  is to determine an appropriate Gliozzi-Scherk-Olive (GSO) projection \cite{Gliozzi:1976qd} of the spectrum. 

To illustrate the NSR formalism we consider the twisted part of the $T^4/\mathbb{Z}_M$ partition function in \eqref{pfk3m}.  The differences in the two formalisms concern only the fermionic theta functions in the numerator. We can  directly obtain the partition function in the NSR formalism by using one of the quartic theta identities of Riemann in a suitable form \cite{Mumford:Tata},
\be 
\prod_{i=1}^4 \vartheta_1 (x_i | \tau) =\frac{1}{2} \pqty{\prod_{i=1}^4 \vartheta_1 (y_i | \tau)  - \prod_{i=1}^4 \vartheta_2 (y_i | \tau) + \prod_{i=1}^4 \vartheta_3 (y_i | \tau)  - \prod_{i=1}^4 \vartheta_4 (y_i | \tau) },  \label{riemth}
\ee
where $y_1 = (x_1 + x_2 + x_3 + x_4 ) /2$,  $y_2 = (x_1 + x_2 - x_3 - x_4)/2$,  $y_3 = (x_1 - x_2 + x_3 - x_4 )/2$,  $y_4 = (x_1 - x_2 - x_3 +x_4)/2$ and we follow the standard theta numbering conventions. The fermionic twists in the second line of \eqref{pfk3m} are $k/N$, $k/N$, $k/N + r/M$ and $k/N -  r/M$ --- the twists of the four fermions in the NSR formalism becomes  $2k/N$, $0$,  $r/M$ and $-r/M$, exactly the same twists as the bosons,  \eqref{zmbos1},  as expected from the NSR formalism by superconformal invariance. 
Various relative signs in \eqref{riemth} enforce the GSO projection with appropriate insertions of the projection operator. 

In the NSR formalism, the SCFT corresponding to the Rindler plane and the SCFT corresponding to the transverse directions are decoupled to begin with.  The two get correlated only through the GSO projection. For toroidal compactifications the internal SCFT is simply the SCFT corresponding to the compact torus such as $T^4$ or $T^6$. In this case, there are no new tachyons besides the ones coming from the dimensional reduction of the ten-dimensional theory. 

For the orbifolds considered here,  we replace the SCFTs corresponding to $T^4$ or $T^6$ by the respective orbifold SCFTs.  We have seen that new tachyons can arise in the doubly twisted sectors. 
For a smooth Calabi-Yau manifold, there are no tachyons to begin with before taking the $\mathbb{Z}_N$ orbifold and there are no doubly-twisted sectors. Therefore, we do not expect new tachyons in addition to the ten-dimensional ones.    It would be interesting to explore further more complicated orbifolds and more general compactifications.   {In this context,  it is worth mentioning that it is possible to extend our method to horizons of other topologies.  In \cite{Dabholkar:2023tzd},  the orbifold construction has been shown to extend to  strings in an an exact black hole background,  namely $\mathrm{AdS}_3 \times S^3 \times T^4$, where the black hole horizon has the topology $S^1 \times S^3 \times T^4$. }

{It is also worth pondering over the questions of other IR aspects involved in the problem.   As mentioned in the discussion following \eqref{modint},  we have worked in the limit of large transverse area. It would be interesting to consider a systematic expansion in area --- the logarithmic term in this expansion is known to encode universal features of the theory.  The set-up of \cite{Dabholkar:2023tzd} is particularly well-suited for addressing such questions.  We hope to return to these questions in future works.}

\acknowledgments

We would like to thank Jeffrey A. Harvey for useful discussions.  UM thanks the following organizations for their hospitality during the progress of this work: Simons Center for Geometry and Physics,  Stony Brook University (during the 2023 Simons Physics Summer Workshop),  the Institute for Advanced Study,  Princeton (during PiTP 2023),   the University of Wisconsin-Madison and ETH Z\"urich.

\appendix

\bibliographystyle{JHEP}
\bibliography{orb-references}

\end{document}